\documentclass[11pt]{article}

\usepackage{amsmath}
\usepackage{amsfonts}
\usepackage{amssymb}
\usepackage{amscd}
\usepackage{latexsym}
\usepackage{mathrsfs}
\usepackage{amsthm}

\newtheorem{Lemma}{Lemma}[section]

\newtheorem{Theorem}[Lemma]{Theorem}

\newtheorem{Remark}[Lemma]{Remark}

\def\B{\mathcal{B}}
\def\C{\mathbb{C}}
\def\CC{\mathscr C}
\def\de{\mathrm d}
\def\D{\Delta}
\def\e{\varepsilon}
\def\EE{\mathscr E}
\def\H{\mathcal{H}}
\def\JJ{\mathscr J}
\def\K{\mathcal{K}}
\def\KK{\mathscr K}
\def\O{\Omega}
\def\Pinf{\mathcal{P}_{\!\infty}}
\def\R{\mathbb{R}}
\def\S{\mathbb{S}}
\def\U{\mathcal{U}}
\def\v{\varphi}
\def\tr{\mbox{\rm tr}}
\def\Tr{\mbox{\rm Tr}}

\begin{document}

\title{A topological version of Levinson's theorem}

\author{Johannes Kellendonk\,~and\,~Serge Richard}
  \date{\small
    \begin{quote}
      \emph{
    \begin{itemize}
    \item[]
			Institut Camille Jordan,
			B\^atiment Braconnier,
			Universit\'e Claude Bernard Lyon 1, \\
			43 avenue du 11 novembre 1918,
			69622 Villeurbanne cedex, France
    \item[]
      \emph{E-mails\:\!:}
      kellendonk@igd.univ-lyon1.fr\,~and\,~srichard@igd.univ-lyon1.fr
    \end{itemize}
      }
    \end{quote}
    April 2005
  }

\maketitle

\begin{abstract}
In the framework of scattering theory, we show how
the scattering matrix can be related to the projection on the 
bound states by an index map of $K$-theory.
Pairings with appropriate cyclic cocyles lead
naturally to a topological version of Levinson's theorem.
\end{abstract}

\section{Introduction}\label{secintro}

Let us consider the self-adjoint operators $H_0:= -\D$ and $H:= H_0 + V$
in the Hilbert space $\H := L^2(\R^n)$, where $|V(x)| \leq
c\;\!(1+|x|)^{-\beta}$  
with $\beta> 1$.
It is well known that for such short range potentials $V$, the wave operators 
\begin{equation}\label{oponde}
\O_{\pm}:=s-\lim_{t \to \pm \infty} e^{itH}\;\!e^{-itH_0}
\end{equation}
exist and have same range.
The complement of this range is generated by the eigenvectors of $H$,
we let $P$ denote the projection on this subspace. 
The scattering matrix $S$ for this system is defined by the product
$\O_+^* \O_-$, where $\O_+^*$ is the adjoint of $\O_+$. 

Levinson's theorem establishes a relation between an expression in terms of the 
unitary operator $S$ and an expression depending on the projection $P$.
There exist many presentations of this theorem, but we recall only the one 
of \cite{Martin} in the case $n=3$. 
We refer to \cite{Osborn}, \cite{Dreyfus} and \cite{Newton} for other versions 
of a similar result.

Let $\U : \H \to L^2\big(\R_+; L^2(\S^{n-1})\big)$ be the unitary
transformation that diagonalizes $H_0$, {\it i.e.}~that satisfies
$[\U H_0 f](\lambda, \omega) = \lambda [\U f](\lambda,\omega)$, with 
$f$ in the domain of $H_0$, $\lambda \in \R_+$ and $\omega \in \S^{n-1}$.
Since the operator $S$ commutes with $H_0$, there exists a family 
$\{S(\lambda)\}_{\lambda \in \R_+}$ of unitary operators in $L^2(\S^{n-1})$
satisfying  $\U S\;\! \U^* = \{S(\lambda)\}$ almost everywhere 
in $\lambda$ \cite[Chap.~5.7]{AJS}.
Under suitable hypotheses on $V$ \cite{Martin} Levinson's theorem takes the 
form 
\begin{equation}\label{LevMartin}
\int_0^\infty \de \lambda \;\big\{\tr[iS(\lambda)^*\hbox{$\frac{\de
    S}{\de \lambda} 
(\lambda)$}]- \hbox{$\frac{\nu}{\sqrt{\lambda}}$}\big\} = 2\pi \;\!\Tr[P],
\end{equation}
where $\tr$ is the trace on $L^2(\S^{n-1})$, 
$\Tr$ the trace on $\H$ and $\nu=(4\pi)^{-1}\int_{\R^3} \de x\;\!V(x)$.
Clearly the r.h.s.\ of this equality is invariant 
under variations of $V$ that do not change the number of bound 
states of $H$.
But it is not at all clear how this stability comes about in the 
l.h.s.

In this note we propose a modification of the l.h.s.\ of \eqref{LevMartin}
in order to restore the topological nature of this equality. The idea
is very natural from the point of view of non-commutative topology: we
rewrite the l.h.s.\ of \eqref{LevMartin} as the result of a pairing
between $K$-theory and cyclic cohomology.  
Beyond formula \eqref{LevMartin}, we show that the unitary $S$ is related 
to the projection $P$ at the level of $K$-theory by the index map,
{\it cf.}~Theorem \ref{thm1}. 
Let us point out that the wave operators play a key role in this work.
Sufficient conditions on $\O_-$ imply that $H$ has only a finite set of
bound states, but also give informations on the behaviour of $S(\cdot)$
at the origin.

\section{The algebraic framework}

In this section we show how the scattering matrix $S$ can be related to the 
projection $P$ on the bound states via a boundary map of $K$-theory. 
Consider the short exact sequence
\begin{equation}\label{sexact}
0 \to C_0\big(\R;\KK\big)\!\rtimes_\tau\!\R\to 
C_0\big(\R\cup\{+\infty\};\KK\big)\!\rtimes_\tau\!\R
\stackrel{ev_{\infty}}{\to} \KK\!\!\rtimes\!\R \to 0,
\end{equation}
where $\KK$ is the algebra 
of compact operators in some Hilbert space. 
The sequence \eqref{sexact} is the Wiener-Hopf extension of the 
crossed product $\KK\!\!\rtimes\!\R$ with trivial $\R$-action on
$\KK$; $\tau$ is the action on
$C_0\big(\R\cup\{+\infty\}\big)$  
by translation, leaving the point $\{+\infty\}$ invariant, and the
surjection $ev_{\infty}$ is induced by evaluation at $\{+\infty\}$.
Our goal is to identify $P$ as an element in the ideal and
$S$ as an element of the unitisation of the quotient, and to verify
that the boundary map $\mbox{ind}:K_1(\KK\!\!\rtimes\!\R)\to K_0\big(
C_0\big(\R;\KK\big)\!\rtimes_\tau\!\R\big)$ maps the $K_1$-class of $S$ to
(minus) the $K_0$-class of $P$. To do so we represent the above
short exact sequence in the physical Hilbert space $\H$. 

Following the developments of \cite{Georgescu} we first
consider the case  $\KK=\C$ and let $A,B$ be (unbounded) self-adjoint
operators in $\H$ both with purely absolutely continuous spectrum
equal to $\R$ 
and commutator given formally by $[iA,B]=-1$. We can then represent 
$C_0\big(\R\cup\{+\infty\};\KK\big)\!\rtimes_\tau\!\R$
faithfully as 
the norm closure $\CC'$ in $\B(\H)$ of the set of finite sums of the form
$\v_1(A) \;\!\eta_1(B)  + \ldots + \v_m(A) \;\!\eta_m(B)$ 
where $\v_i\in C_0\big(\R\cup\{+\infty\}\big)$ and
$\eta_i\in C_0(\R)$. We denote by $\JJ'$ the ideal obtained by
choosing functions $\v_i$ that vanish at $\{+\infty\}$.
Furthermore, we can represent $\KK\!\!\rtimes\!\R$ faithfully
in $\B(\H)$ by elements of the form $\eta(B)$ with $\eta\in C_0(\R)$. 
This algebra is denoted by $\EE'$. 

In \cite{Georgescu} position and momentum operators were chosen for
$A$ and $B$ but we take $A := -\hbox{$\frac{i}{2}$}(Q\cdot \nabla
+ \nabla \cdot Q)$ and $B:= \hbox{$\frac{1}{2}$}\ln H_0$.
We refer to \cite{Jensen} for a thorough description of $A$ 
in various representations.
Let us notice that a typical element of $\CC'$ is of the form
$\v(A)\;\!\eta(H_0)$ with $\v \in C_0\big(\R\cup\{+\infty\}\big)$
and $\eta \in C_0(\R_+)$, the algebra of continuous functions on $\R_+$ 
that vanish at the origin and at infinity.
We shall now consider $\KK=\K\big(L^2(\S^{n-1})\big)$ from the decomposition 
$\H \cong L^2\big(\R_+;L^2(\S^{n-1})\big)$ in spherical coordinates. 
Since $A$ and $H_0$ are rotation invariant the presence of a larger $\KK$ 
does not interfere with the above argument.
Thus we set $\CC:= \CC'\otimes\KK$, $\JJ:=\CC'\otimes \KK$ and
$\EE:=\EE'\otimes \KK$. These algebras are all
represented in the same Hilbert space $\H$, although $\EE$ is a
quotient of $\CC$. 
The surjection $ev_{\infty}$ becomes the map
$\Pinf$, where
$\Pinf[T]:= T_\infty$, with $T_\infty$ uniquely defined by 
the conditions $\|\chi(A\geq t)\;\!(T-T_\infty)\|\to 0$ and 
$\|\chi(A\geq t)\;\!(T^*-T^*_\infty)\|\to 0$ as $t \to +\infty$, 
$\chi$ denoting the characteristic function.
We easily observe that $\Pinf[\v(A)\;\!\eta(H_0)] = \v(+\infty)\;\!\eta(H_0)$
for any $\v \in C_0\big(\R\cup\{+\infty\}\big)$ and 
$\eta \in C_0\big(\R_+; \KK\big)$, where $\v(+\infty)$ is simply the value
of the function $\v$ at the point $\{+\infty\}$. 
Let us summarise our findings: 

\begin{Lemma}\label{repdesalg}
All three algebras of \eqref{sexact} are represented faithfully in
$\H$ by $\JJ$, $\CC$ and $\EE$. In $\B(\H)$ the surjection
$ev_{\infty}$ becomes $\Pinf$.
\end{Lemma} 

Note that $\JJ$ is equal to the set 
of compact operators in $\H$.
For suitable potentials $V$, the operator $S-1$ belongs 
to $\EE$ \cite{Jensen,JK} and $P$ is a compact operator. 
The key ingredient below is the use of $\O_-$ to make the link between 
the $K_1$-class $[S]_1$ of $S$ and the $K_0$-class $[P]_0$ of $P$.

\begin{Theorem}\label{thm1}
Assume that $\O_- -1$ belongs to $\CC$. Then $S-1$ is an element of $\EE$, 
$P$ belongs to $\JJ$ and one has at the level of $K$-theory:
\begin{equation}\label{Kegalite}
\mbox{\rm ind} [S]_1 = -\; [P]_0.
\end{equation}
\end{Theorem}

\begin{proof}
Let $T\in \CC$. Then  $T_\infty=\Pinf(T) \in \EE$ satisfies
$\|\chi(A\geq t)(T-T_\infty)\|\to 0$ as $t \to +\infty$. Equivalently, 
$\|\chi(A\geq 0)\;\![U(t) T U(t)^* -T_\infty]\| \to 0$ as $t \to +\infty$, since
$T_\infty$ commutes with $U(t):=e^{\frac{i}{2}t\ln H_0}$ for all $t \in \R$.
It is then easily observed that 
$s-\lim_{t\to +\infty}U(t)\;\!T\;\!U(t)^* = T_\infty$.
Now, if $T$ is replaced by $\O_- -1$, the operator $T_\infty$ has to be equal to
$S-1$, since $s-\lim_{t \to +\infty} U(t)\;\!\O_-\;\!U(t)^*$ is equal to $S$.
Indeed, this result directly follows from the intertwining relation of $\O_-$ 
and the invariance principle \cite[Thm.~7.1.4]{ABG}.

We thus have shown that $\O_--1$  is a preimage of $S-1$ in $\CC$. 
It is well known that $\O_- \O_-^* = 1-P$ and $\O_-^* \O_- = 1$. In
particular $\O_-$ is a partial isometry so that
$\mbox{\rm ind}[S]_1 = [ \O_- \O_-^*]_0-[\O_-^* \O_-]_0=-[P]_0$, see 
{\it e.g.}~\cite[Prop.~9.2.2]{RLL}.
\end{proof}

\begin{Remark}
{\rm It seems interesting that the condition $\O_- -1 \in \CC$ implies
the finiteness of the set of eigenvalues of $H$. Another consequence of this 
hypothesis is that $S(0)=1$, a result which is also not obvious. See
\cite[Sec.~5]{JK} 
for a detailed analysis of the behaviour of $S(\cdot)$ near the origin.}
\end{Remark}

It is important to express the above condition on $\O_-$ in a more
traditional way, {\it i.e.}~in terms of scattering conditions.
The following lemma is based on an alternative description of the 
$C^*$-algebra $\CC$. Its easy proof can be obtained by mimicking
some developments given in Section 3.5 of \cite{Georgescu}.
We also use the convention of that reference, that is: if a symbol
like $T^{(*)}$ appears in a relation, it means that this relation has to hold
for $T$ and for its adjoint $T^*$.

\begin{Lemma}\label{appartenance}
The operator $\O_-$ belongs to $\CC$ if and only if the following conditions
are satisfied:
\begin{enumerate}
\item[{\rm (i)}] $\|\chi(H_0\leq \e) \;\!(\O_- -1)^{(*)}\| \to 0$ as $\e \to 0$, and
 $\|\chi(H_0\geq \e) \;\!(\O_- -1)^{(*)}\| \to 0$ as $\e \to +\infty$,
\item[{\rm (ii)}] $\|\chi(A \leq t) \;\!(\O_- -1)^{(*)}\| \to 0$ as $t \to -\infty$, and
 $\|\chi(A \geq t) \;\!(\O_- -S)^{(*)}\| \to 0$ as $t \to +\infty$.
\end{enumerate}
Equivalently, the condition {\rm (ii)} can be rewritten as
\begin{enumerate}
\item[{\rm (ii')}]  $\|\chi(A \leq 0) \;\!U(t)\;\!(\O_- -1)^{(*)}\;\!U(t)^*\| \to 0$ as 
$t \to -\infty$, and $\|\chi(A \geq 0) \;\!U(t)\;\!(\O_- -S)^{(*)}\;\!U(t)^*\| 
\to 0$ as $t \to +\infty$.
\end{enumerate}
\end{Lemma}

\section{A new version of Levinson's theorem}

In the next statement, it is required that the map 
$\R_+ \ni \lambda \mapsto S(\lambda) \in \B\big(L^2(\S^{n-1})\big)$ is 
differentiable. 
We refer for example to \cite[Thm.~3.6]{Jensen} for sufficient conditions 
on $V$ for that purpose.
Trace class conditions on $S(\lambda)-1$ for all $\lambda \in \R_+$ are common requirements
\cite{Davies}. Unfortunately, similar conditions on $S'(\lambda)$ were
much less studied in the literature.

\begin{Theorem}\label{Levnous}
Let $\O_- -1$ belong to $\CC$. Assume furthermore 
that the map
$\R_+\ni \lambda \mapsto S(\lambda) \in \B\big(L^2(\S^{n-1})\big)$ is 
differentiable, and that $\lambda\mapsto \tr [S'(\lambda)]$
belongs to $L^1\big(\R_+, \de \lambda\big)$.
Then the following equality holds:
\begin{equation}\label{new}
\int_0^\infty \de \lambda \;\tr\big[i(S(\lambda)-1)^*\;\!S'(\lambda)\big] = 
2\pi \;\!\Tr[P].
\end{equation}
\end{Theorem}

\begin{proof}
The boundary maps in $K$-theory of the exact sequence \eqref{sexact}
are the inverses of the Connes-Thom isomorphism (which here
specialises to the Bott-isomorphism as the action in the quotient is
trivial) and have a dual in
cyclic cohomology \cite{Connes},
or rather on higher traces \cite{Connes2,Johannes}, which gives rise
to an equality between pairings which we
first recall: 
$\Tr$ is a $0$-trace on
the ideal $C_0\big(\R;\KK\big)\!\rtimes_\tau\!\R\cong
\K\big(L^2(\R)\big)\otimes \K\big(L^2(\S^{n-1})\big)$ which we factor
$\Tr=\Tr'\otimes \tr$. Then $\hat \tr:  \KK\!\!\rtimes\!\R \to \C$,
$\hat \tr [a] = \tr[a(0)]$ is a trace on the crossed product and
$(a,b) \mapsto \hat \tr [a \delta(b)]$ a $1$-trace where
$[\delta(b)](t) = it b(t)$. With these ingredients 
\begin{equation}\label{pair} 
\hat \tr [i(u-1)^* \delta(u)] = -2\pi\Tr [p]
\quad\mbox{if}\quad \mbox{\rm ind}[u]_1 = [p]_0 ,
\end{equation}
provided $u$ is a representative of its $K_1$-class $[u]_1$ on which
the $1$-trace can be evaluated. This is for instance the case if
$\delta(u)$ is $\hat\tr$-traceclass.
To apply this to our situation, in which $u$ is the unitary
represented by the scattering matrix and $p$ is represented by
the projection onto the bound states, we express $\delta$ and $\hat
tr$ on $\U\EE\U^*$ where $\U$ is the unitary from Section~1
diagonalising $H_0$. Then $\delta$ becomes $\lambda\frac{\de}{\de\lambda}$
and $\hat\tr$ becomes $\int_{\R_+} \hbox{$\frac{\de
    \lambda}{\lambda}$} \;\!\tr$. Our 
hypothesis implies the neccessary trace class property so that
the l.h.s.~of (\ref{pair}) corresponds to 
$ \int_0^\infty \de \lambda \;\!\tr\big[i(S(\lambda)-1)^*
S'(\lambda)\big]$ and the r.h.s.~to $2\pi\;\!\Tr[P]$.
\end{proof}

\begin{Remark}
{\rm Expressions very similar to
\eqref{new} already appeared in \cite{Osborn} and \cite{Dreyfus}.
However, it seems that they did not attract the attention of the
respective authors and that a formulation closer to
\eqref{LevMartin} was preferred.  One reason is that the operator
$\{S(\lambda)^*S'(\lambda)\}_{\lambda \in \R_+}$ has a physical
meaning: it represents the {\em time delay} of the system under
consideration.  We refer to \cite{AC} for more explanations and
results on this operator.}
\end{Remark}

\begin{Remark}
{\rm At present our approach does not allow to say anything about a
{\em half-bound state}. We refer to \cite{JK}, \cite{Newton} or
\cite{Newton2}   
for explanations on that concept and to \cite{Newton} or \cite{Newton2} for
corrections of Levinson's theorem in the presence of such a 
{\em 0-energy resonance}.}
\end{Remark}

\section{Further prospects}

We outline several improvements or extensions that ought to be carried
out or seem natural in view of this note. We hope to express some of these
in a further publication.
\begin{itemize}
\item 
Our main hypothesis of Theorem~\ref{Levnous}, that $\O_--1$
belongs to the $C^*$-algebra $\CC$, is crucial and we have provided
estimates in Lemma \ref{appartenance} which would guarantee it.
Such estimates are rather difficult to obtain and we were
not able to locate similar conditions in the literature. They clearly
need to be addressed.
\item Similar results should hold for a more general
operator $H_0$ with absolutely continuous spectrum.
In that case, the role of $A$ would be played by an operator
conjugate to $H_0$. We refer to \cite[Prop.~7.2.14]{ABG} for the 
construction of such an operator in a general framework.
\item
More general short range potentials or trace class perturbations can also 
be treated in a very similar way. By our initial hypothesis on $V$ we have
purposely eliminated positive eigenvalues of $H$, but it would be interesting
to have a better understanding of their role with respect to 
Theorems \ref{thm1} and \ref{Levnous}.
\item 
In principle, Theorem~\ref{thm1} is stronger than Levinson's
theorem and one could therefore expect new topological
relations from pairings with other cyclic cocycles. In the present
setting these do not yet show up as the ranks of the
$K$-groups are too small. But in more complicated scattering
processes this could well be the case.    
\item 
In the literature one finds also the
so-called {\it higher-order Levinson's Theorems} \cite{Bolle}. 
In the case $n=3$ and under suitable hypotheses they take the form
\cite[eq.~3.28]{Bolle} 
\begin{equation*}
\int_0^\infty \de \lambda \;\!\lambda^{\!\hbox{\tiny \it N}}
\big\{\tr \big[iS(\lambda)^*S'(\lambda)\big] - C_{\!\hbox{\tiny \it
    N}}(\lambda)\big\} = 2\pi \sum_j e_j^{\hbox{\tiny \it N}}, 
\end{equation*}
where $N$ is any natural number, $C_{\!\hbox{\tiny \it N}}$ are correction terms, 
and $\{e_j\}$ is the set of eigenvalues of $H$ with multiplicities counted. 
The correction terms can be explicitly computed in terms of 
$H_0$ and $V$ \cite{Bolle} and we expect that they can be absorbed in
a similar manner into the $S$-matrix as above. 
\end{itemize}

\section*{Acknowledgements}
Serge Richard thanks the Swiss National Science Foundation for
its financial support.


\begin{thebibliography}{00}

\bibitem{ABG} W.O. Amrein, A. Boutet de Monvel, V. Georgescu, $C_0$-groups,
commutator methods and spectral theory of N-body Hamiltonians, Progress
in Math. vol.~135, Birkh\"auser, 1996.

\bibitem{AC} W.O. Amrein, M.B. Cibils, Global and Eisenbud-Wigner time
delay in scattering theory, Helv. Phys. Acta 60 (1987) 481--500.

\bibitem{AJS} W.O. Amrein, J.M. Jauch, K.B. Sinha, Scattering theory
in quantum mecanics, W.A. Benjamin, 1977.

\bibitem{Bolle} D. Boll\'e, Higher-order Levinson's theorems and the
high-temperature expansion of the partition function, Ann. Phys. 121 
(1979) 131--146.

\bibitem{Osborn} D. Boll\'e, T.A. Osborn, An extended Levinson's
theorem, J. Math. Phys. 18 (1977) 432--440.

\bibitem{Connes} A. Connes, An analogue of the Thom isomorphism for
crossed products of a $C^*$-algebra by an action of $\R$, 
Adv. in Math. 39 (1981) 31--55. 

\bibitem{Connes2}
A. Connes, Cyclic cohomology and the transverse fundamental class
of a foliation, in: Geometric methods in operator algebras, Kyoto,
1983, pp.~52--144, Pitman Res.~Notes in Math., Longman, Harlow, 1986. 

\bibitem{Davies} E.B. Davies, Energy dependence of the scattering operator,
Adv. App. Math. 1 (1980) 300--323.

\bibitem{Dreyfus} T. Dreyfus, The determinant of the scattering matrix
and its relation to the number of eigenvalues, J. Math. Anal. Appl. 64 
(1978) 114--134, {\it and} 
The number of states bound by non-central potentials, Helv. Phys. Acta 51 
(1978) 321--329.

\bibitem{Georgescu} V. Georgescu, A. Iftimovici, $C^*$-algebras of
quantum Hamiltonians, in: Operator Algebras and Mathematical Physics, Conference
Proceedings: Constan\c{t}a (Romania) July 2001, pp.~123--167, Theta Foundation, 2003.

\bibitem{Jensen} A. Jensen, Time-delay in potential scattering theory,
Commun. Math. Phys. 82 (1981) 435--456.

\bibitem{JK} A. Jensen, T. Kato, Spectral properties of Schr\"odinger
operators and time-decay of the wave functions, Duke Math. J. 46 (1979) 583--611.

\bibitem{Johannes} J. Kellendonk, H. Schultz-Baldes, Boundary maps for
$C^*$-crossed product with $\R$ with an application to the quantum Hall effect,
Commun. Math. Phys. 249 (2004) 611--637.

\bibitem{Martin} Ph.A. Martin, Time delay of quantum scattering
processes, Acta Phys. Austriaca, Suppl. XXIII (1981) 157--208.

\bibitem{Newton} R.G. Newton, Noncentral potentials: The generalized
Levinson theorem and the structure of the spectrum, J. Math. Phys. 18 (1977) 
1348--1357, {\it and} 
Nonlocal interactions: The generalized Levinson theorem and the
structure of the spectrum, J. Math. Phys. 18 (1977) 1582--1588.

\bibitem{Newton2} R.G. Newton, The spectrum of the Schr\"odinger $S$
matrix: Low energies and a new Levinson theorem, Ann. Phys 194 (1989) 173--196.

\bibitem{RLL} M. Rordam, F. Larsen, N.J. Laustsen, An introduction to
$K$-theory for $C^*$-algebras, London Mathematical Society Student Texts 49,
Cambridge University Press, 2000. 

\end{thebibliography}
\end{document}